\documentclass[prl,twocolumn,aps]{revtex4-1}
\usepackage{color}
\usepackage{epsfig,amssymb,amsmath,latexsym,bm}

\newcommand\beq{\begin{equation}}
\newcommand\eeq{\end{equation}}
\newcommand\bea{\begin{eqnarray}}
\newcommand\eea{\end{eqnarray}}

\begin{document}

\title{Geometric Quantum Noise of Spin}

\author{Alexander \surname{Shnirman}$^{1,5}$, Yuval \surname{Gefen}$^{2,3}$, Arijit \surname{Saha}$^4$, Igor S. \surname{Burmistrov}$^{5,6}$, Mikhail N.  \surname{Kiselev}$^7$, Alexander \surname{Altland}$^8$}
\affiliation{$^1$Institut f\"ur Theorie der Kondensierten Materie and DFG-Center for Functional Nanostructures (CFN), Karlsruhe Institute of Technology, D-76128 Karlsruhe, Germany}
\affiliation{$^2$Department of Condensed Matter Physics, Weizmann Institute of Science, 76100 Rehovot, Israel}
\affiliation{$^3$Institut f\"ur Nanotechnologie, Karlsruhe Institute of Technology, 76021 Karlsruhe, Germany}
\affiliation{$^4$Department of Physics, University of Basel, CH-4056 Basel, Switzerland}
\affiliation{$^5$L.D. Landau Institute for Theoretical Physics RAS, Kosygina street 2, 119334 Moscow, Russia}
\affiliation{$^6$Moscow Institute of Physics and Technology, 141700 Moscow, Russia} 
\affiliation{$^7$International Center for Theoretical Physics, Strada Costiera 11, I-34014 Trieste, Italy}
\affiliation{$^8$Institut f\"ur Theoretische Physik, Universit\"at zu K\"oln, Z\"ulpicher Str. 77, D-50937 K\"oln, Germany}

\date{\today}


\begin{abstract}
The presence of geometric phases is known to affect the dynamics of the systems involved. Here we consider a quantum degree of freedom, moving in a dissipative environment, whose dynamics is described by a Langevin equation with quantum noise. We show that geometric phases enter the stochastic noise terms. Specifically, we 
consider small ferromagnetic particles (nano-magnets) or quantum dots close to Stoner instability, and investigate 
the dynamics of the total magnetization in the presence of tunneling coupling to the metallic leads. 
We generalize the Ambegaokar-Eckern-Sch\"on (AES) effective action and the corresponding semiclassical equations 
of motion from the U(1) case of the charge degree of freedom to the SU(2) case of the magnetization. The Langevin 
forces (torques) in these equations are strongly influenced by the geometric phase. 
As a first but nontrivial application we predict low temperature quantum diffusion of the magnetization 
on the Bloch sphere, which is governed by the geometric phase. We propose a protocol for experimental observation 
of this phenomenon. 
\end{abstract}

\maketitle

{\it Introduction.}
It is well known that the kinetic part of the action of a free spin of length $S$, whose position is described in spherical coordinates by angles $\theta$ and $\phi$ reads ${\cal S}_{spin} = \int p d q$. Here the generalized coordinate is $q \equiv \phi$ and the conjugate momentum is $p \equiv S(1-\cos\theta)$. This action, a.k.a. geometric (Berry) phase action or Wess-Zumino-Novikov-Witten 
(WZNW) action, produces deterministic spin dynamics if accompanied by, e.g., a Zeeman term. If the spin is subject to dissipation its equations of motion are expected to contain deterministic friction terms, e.g., Gilbert damping, as well as stochastic Langevin terms. Here we show that the geometric phase determines the form of these stochastic terms, and analyze the consequence of this for observables. Specifically we focus on the dynamics of the collective spin degree of freedom of either a nano-magnet or a paramagnetic quantum dot near the Stoner instability characterized by a large total spin~\cite{PhysRevB.62.14886,PhysRevLett.96.066805,JETPLetters.92.179,SahaAnnals,PhysRevB.85.155311}. The system is tunnel coupled to a normal lead, which gives rise to a dissipative behavior.  

We find that in the quantum regime, i.e., when the precession frequency is higher than the temperature, the stochastic spin 
torques, represented through random Langevin terms,
are substantially influenced by the Berry phase accumulated by the system in the course of precession. 
As an application of our theory we calculate the diffusion rate for a large spin, which is artificially held on a high-energy precessing trajectory by a specific multiple echo (``bang-bang'') protocol~\cite{BangBang}.  

Our approach can be viewed as a generalization of the Landau-Lifschitz-Gilbert (LLG)-Langevin 
equation~\cite{Gilbert2004,BrownLLGLangevin}, central to the field of spintronics~\cite{RevModPhys.77.1375}, to 
a regime where quantum dynamics dominates. 
Stochastic LLG equations have been derived in numerous publications for both a localized spin 
in an electronic environment (a situation of the Caldeira-Leggett type)~\cite{PhysRevB.73.212501,PhysRevB.85.115440} and for a magnetization formed by itinerant electrons~\cite{ChudnovskiyPRL,BaskoVavilovPRB2009}.
In all these papers the precession frequency was assumed to be lower than the temperature or the voltage, thus 
justifying the semi-classical treatment of the problem. In this regime the geometric phase
did not influence the Langevin terms. 
 
From a different perspective, the equation-of-motion presented here is derived from a new action which constitutes a generalization of the Ambegaokar-Eckern-Sch\"on (AES) theory~\cite{AES_PRL,AES_PRB}. The latter was written to describe the dynamics of the charge degree of freedom (marked by an Abelian U(1) symmetry). 
Our generalized AES action, which is the first main result of our analysis, is underscored by the non-Abelian SU(2) dynamics.   
As only two out of three SU(2) Euler angles are needed to describe the spin position, a gauge freedom emerges. A central 
element of our analysis is to employ this freedom and find a gauge, which allows for efficient calculation and highlights the role of the Berry phase in the stochastic Langevin terms. 

{\it The effective action.}
Our derivation here is technically close to that of Ref.~\cite{ChudnovskiyPRL}. However, in contrast 
to Ref.~\cite{ChudnovskiyPRL}, we do not limit ourselves to small deviations of the spin from the instantaneous 
direction, but rather consider the action on global trajectories covering the whole Bloch sphere.
 
To demonstrate the emergence of an AES-like effective action we consider 
a quantum dot with strong exchange interaction coupled to a normal lead. 
The Hamiltonian reads
$H = H_{dot} + H_{lead} + H_{tun}$.
The quantum dot is described by the magnetic part~\footnote{
Here we disregard the charging part of the ''universal'' Hamiltonian, having in mind, e.g., systems of the type considered in Refs.~\cite{ChudnovskiyPRL,BaskoVavilovPRB2009}. Consequently no Kondo physics is expected. (See also Supplemental Material E.)} 
of the "universal" Hamiltonian~\cite{PhysRevB.62.14886} 
\begin{equation}\label{eq:Hdot}
H_{dot} = \sum_{\alpha,\sigma} \epsilon^{\phantom\dag}_\alpha  a^\dag_{\alpha,\sigma} a^{\phantom \dag}_{\alpha,\sigma} - 
J \bm{S}^2 + \bm{B}\bm{S}\ ,
\end{equation}
where 
$
\bm{S}\equiv (1/2) \sum_{\alpha,\sigma_1,\sigma_2}\,a^\dag_{\alpha,\sigma_1} \bm{\sigma}_{\sigma_1,\sigma_2}\, a^{\phantom \dag}_{\alpha,\sigma_2}
$
is the operator of the total spin on the quantum dot, $\bm{B}$ is the external magnetic field,  and $J>0$ is the corresponding ``zero mode'' ferromagnetic exchange constant. The Hamiltonian of the lead and that 
describing the tunneling between the dot and the lead are standard:
$H_{lead} =  \sum_{\gamma,\sigma} \epsilon^{\phantom\dag}_{\gamma}  c^\dag_{\gamma,\sigma} c^{\phantom \dag}_{\gamma,\sigma}$ and $H_{tun} =  \sum_{\alpha,\gamma,\sigma} V^{\phantom \dag}_{\alpha,\gamma} a^{\dag}_{\alpha,\sigma} c^{\phantom  \dag}_{\gamma,\sigma}  + h.c.$. We assume here a non-magnetic lead. 

We consider the Keldysh generating functional
${\mathcal {Z}} = \int D\bar \Psi D\Psi \,\exp{[i\,{\cal S}_\Psi]}$,  
where the Keldysh action is given by 
${\cal S}_\Psi  =  \oint_K  d t\, (i {\bar\Psi} \partial_t \Psi - H )$ 
(plus the necessary source terms which are not explicitly written).
Here, for brevity, $\Psi$ denotes all fermionic fields and the time $t$ runs along the Keldysh contour. 
After standard Hubbard-Stratonovich manipulations~\cite{KamenevBook,JETPLetters.92.179,SahaAnnals}, decoupling the interaction term 
$- J \bm{S}^2$ we obtain ${\mathcal {Z}} = \int D \bm{\mathcal{M}} \,\exp{[i\,{\cal S}_M]}$ and
the action for the bosonic vector $\bm{\mathcal{M}}(t)$ reads
\begin{equation}\label{SPhi}
i {\cal S}_{M} =  
\mathrm{tr\;ln} \left[\left(
\begin{array}{cc}
G_{dot}^{-1} &  - {\hat V} \\
-{\hat V}^\dag & G_{lead}^{-1} 
\end{array}
\right)\right] - i\, \oint\limits_K d t\, \frac{|\bm{\mathcal{M}}|^2}{4J}\ .
\end{equation}
Here 
$G_{dot}^{-1}  \equiv [ i\partial_t - \epsilon_\alpha  -
(\bm{\mathcal{M}}(t)+\bm{B})\cdot\bm{\sigma}/2]$, while 
$G_{lead}^{-1}  \equiv  i\partial_t - \epsilon_{\gamma}$. 
Both $G_{dot}^{-1}$ and $G_{lead}^{-1}$ are matrices with time, spin, and orbital indexes. 
We introduce $\bm{M}(t) \equiv \bm{\mathcal{M}}(t) + \bm{B}$.
Expanding (\ref{SPhi}) in powers of the tunneling matrix $\hat V$ and re-summing we easily obtain
\begin{equation}
i {\cal S}_{M} =  \mathrm{tr\;ln} \left[G_{lead}^{-1} \right] +  \mathrm{tr\;ln} \left[G_{dot}^{-1} - \Sigma\right] 
- i\, \oint\limits_K d t\, \frac{|\bm{M}-\bm{B}|^2}{4J}\ ,
\end{equation}
where the self energy reads
$\Sigma \equiv {\hat V} G_{lead} {\hat V}^\dag$. The first term is trivial, i.e., it would never contain the source fields. Thus, it will be dropped in what follows. 

{\it Rotating frame.}
We introduce a unit length vector $\bm{n}(t) = (\sin\theta\cos\phi, \sin\theta\sin\phi,\cos\theta)$ through
$\bm{M}(t) = M(t)\bm{n}(t)$ and transform to a coordinate system in which $\bm{n}$ coincides with the $z$-axis 
$\bm{n}(t)\cdot\bm{\sigma} =  R(t) \sigma_z R^{\dag}(t)$. This condition 
identifies the unitary rotation matrix $R$ as an element of ${\rm SU(2)/U(1)}$. Indeed, if we employ the Euler angle representation
$
R = \exp{\left[-(i\phi/2) \sigma_z\right]} \exp{\left[-(i\theta/ 2) \sigma_y\right]} \exp{\left[-(i\psi/2) \sigma_z\right]}
$, 
then the angles $\phi(t)$ and $\theta(t)$ determine the direction of $\bm{n}(t)$, while $\psi(t)$ is arbitrary, i.e., 
the condition $\bm{n}(t)\cdot\bm{\sigma} =  R \sigma_z R^{\dag}$ is achieved with any value of $\psi(t)$. 
Thus, $\psi$ represents the gauge freedom of the problem. We introduce, first, a shifted gauge field $\chi(t) \equiv  
\phi(t) + \psi(t)$.
This way a periodic boundary condition, e.g., in the Matsubara 
representation $R(\tau) = R(\tau + \beta)$, is satisfied for $\chi(\tau+\beta)=\chi(\tau)+4\pi m$ (The fact that $m$ is integer is intimately related to the spin quantization~\cite{AbanovAbanov}). 
We can always assume trivial boundary conditions for $\chi$, i.e., $m=0$. 
We keep this representation of the rotation matrix $R$ also for the Keldysh technique. 

We perform a transition to the rotating frame and obtain
$i{\cal S}_{M} =  \mathrm{tr\;ln} \left[R^{\dag}\left(G_{dot}^{-1} - \Sigma \right)R\right] 
-i\, 
\oint_K d t\, ({M^2} -2 \bm{B}\bm{M})/(4J)$ 
(we omit the constant term $\propto |\bm{B}|^2$).
For the Green's function of the dot this gives 
$
R^{\dag}G_{dot}^{-1} R =  i\partial_t - \epsilon_\alpha  - 
M(t)\,\sigma_z/2 - Q$, 
where we define the gauge (Berry) term as
$
Q \equiv R^{\dag}(-i \partial_t ) R 
= Q_\parallel + Q_\perp$. 
Here
$
Q_\parallel \equiv 
[\dot \phi(1-\cos\theta) - \dot\chi]\,\sigma_z/2 
$
and
$
Q_\perp \equiv 
-
\exp{\left[i\chi\sigma_z\right]}
\left[\dot\theta\,\sigma_y - \dot \phi \sin\theta \,\sigma_x  \right]\,\exp{\left[i\phi\sigma_z\right]}/2
$. 
Note, that $Q$ depends on the choice of the gauge field $\chi$. Finally, we obtain
\begin{eqnarray}
i{\cal S}_{M} &=&\mathrm{tr\;ln} \left[G_{dot,z}^{-1} -Q - R^{\dag}\Sigma R \right] \nonumber\\
&-&i\, \oint\limits_K d t \left[\frac{M^2}{4J} -\frac{\bm{B}\bm{M}}{2J}\right]\ ,
\end{eqnarray}
where $G_{dot,z}^{-1} \equiv  i\partial_t - \epsilon_\alpha - 
(1/2)\,M(t)\,\sigma_z$.

To find the semi-classical trajectories of the magnetization we need to consider paths 
$M(t)$, $\theta(t)$, $\phi(t)$ on the Keldysh contour such that the quantum components are small 
(in Supplemental Material C we discuss the physical meaning of this approximation).  
The quantum ($q$) and classical ($c$) components of the fields are expressed in terms of the forward  ($u$) and backward ($d$) 
components~\cite{KamenevBook}, e.g.,  $\phi_{q}(t)=\phi_u(t)-\phi_d(t)$ and $\phi_{c}(t)=(\phi_u(t)+\phi_d(t))/2$.
Performing the standard rotation~\cite{KamenevBook} we thus obtain 
\begin{eqnarray}\label{SPhiRotated}
i {\cal S}_{M} &=&\mathrm{tr\;ln} \left[\tilde G_{dot,z}^{-1}  - \tilde Q  - \tilde R^\dag\tilde \Sigma\tilde R \right] 
\nonumber\\
&+& i\,\int d t \,\frac{\bm{B} \bm{M}_q}{2J}- i\,\int d t \,\frac{M_cM_q}{2J}\ ,
\end{eqnarray}
where
$\tilde G_{dot,z}^{-1} \equiv \tau_x G_{dot,z}^{-1}$. 
The local in time matrix fields $Q(t)$ and $R(t)$ also acquire the $2\times 2$ matrix structure in the Keldysh space, 
e.g., $\tilde Q = Q_c \tau_x + Q_q \tau_0 /2$, where $\tau_{x,y,z,0}$ are the standard Pauli matrices.

{\it The adiabatic limit.}
Thus far we have made no approximations. The action (\ref{SPhiRotated}) governs both the dynamics of the magnetization 
amplitude $M(t)$ and of the magnetization direction $\bm{n}(t)$.
Here we focus on the case of a large amplitude $M$ (more precisely, $M$ fluctuates around a large average value $M_0$ (see also Supplemental Material F)). Such a situation arises either on the ferromagnetic side of the Stoner transition or on the paramagnetic side, but very close to the transition. In the latter case, as was shown in Refs.~\cite{JETPLetters.92.179,SahaAnnals}, it is the 
integration out of the fast angular motion of $\bm{n}$ which creates an effective potential for $M$, forcing it to acquire a finite average value. More precisely the angular motion with frequencies $\omega  \gg \max{[T,B]}$ (we
adopt the units $\hbar=k_B=1$) can be integrated out, renormalizing the effective potential for the slow part of $M(t)$. The very interesting question of the dissipative dynamics of slow longitudinal fluctuations of $M(t)$ in the mesoscopic Stoner regime will be addressed elsewhere. 
Here we focus on the slow angular motion and substitute $M(t)=M_0$. Thus, the last term of (\ref{SPhiRotated}) can be dropped. We note that in the adiabatic limit we may neglect $\tilde Q_\perp$ as it contributes only in the second order in
$d\bm{n}/dt$~\cite{SahaAnnals}.

The idea now is to expand the action (\ref{SPhiRotated}) in both $\tilde Q$ (which is small due to the slowness of $\bm{n}(t)$) 
and $\tilde R^\dag\tilde \Sigma\tilde R$ (which is small due to the smallness of the tunneling amplitudes). 
A straightforward analysis reveals that a naive expansion 
to the lowest order in both violates the gauge invariance with respect to the choice of $\chi(t)$. 
One can show that the expansion in $\tilde R^\dag\tilde \Sigma\tilde R$ is gauge invariant only if all orders 
of $\tilde Q$ are taken into account, that is if $(\tilde G_{dot,z}^{-1}  - \tilde Q)^{-1}$ is used as zeroth order Green's function 
in the expansion. This problem necessitates a clever choice of gauge, such that 
$(\tilde G_{dot,z}^{-1}  - \tilde Q)^{-1}$ is as close as possible to $\tilde G_{dot,z}$, i.e., the effect of $\tilde Q$ is ``minimized''.

{\it Choice of gauge.}
As the action (\ref{SPhiRotated}) is gauge invariant we are allowed to choose the most convenient form of $\chi(t)$. 
We make the following choice
\begin{equation}
\begin{split}
\dot \chi_c(t) & = \dot \phi_c(t)  \,(1-\cos\theta_c(t)) \ , \\
\chi_q(t) & = \phi_q(t)\,(1-\cos\theta_c(t))\ ,
\end{split}
\label{KeldyshGaugeFix}
\end{equation}
which satisfies the necessary boundary conditions, i.e., $\chi_q(t=\pm\infty)=0$.

We next motivate the choice of~Eq. (\ref{KeldyshGaugeFix}). Ideally we should have chosen a gauge that would 
lead to $Q_\|=0$. However, any gauge has to satisfy the boundary condition
$\chi_q(t=\pm \infty) = 0$. This condition is violated by the naive gauge, in which on both forward and backward 
Keldysh contours $\dot \chi = \dot \phi (1-\cos \theta)$, and, thus, $Q_\|$ 
vanishes identically.  The gauge (\ref{KeldyshGaugeFix}) 
satisfies the boundary conditions and leads to the desired cancellation $Q_{\|,c}=0$, whereas the quantum 
component of $Q_\|$ remains nonzero:
\begin{equation}
\label{Qparq}
Q_{\parallel,q} =\frac{1}{2}\, \sigma_z\, \sin\theta_c \,\left[\dot \phi_c \theta_q  - \dot \theta_c \phi_q\right]\ .
\end{equation}
At the same time this choice allows for the expansion of the Keldysh action in the small $\phi_q$ and $\theta_q$ as 
there are no $\dot \phi_q$ terms in (\ref{Qparq}) (see Supplemental Material A).

{\it Berry phase (WZNW action).}
Expanding the zeroth order in $\tilde \Sigma$ term of the action (\ref{SPhiRotated}) to first order in $\tilde Q$ we obtain the 
well known in spin physics (see, e.g., Refs.~\cite{Volovik87,AbanovAbanov}) Berry phase (WZNW) action
$i{\cal S}_{WZNW} =
-\frac{1}{2}\, \int dt\, \mathrm{tr} \left[ G^K_{dot,z}(t,t) Q_{\parallel,q}(t)\right]$, which after a straightforward calculation reads
\begin{eqnarray}
\label{WZNW}
i{\cal S}_{WZNW} =i S \,\int dt \,\sin\theta_c \,\left[\dot \phi_c \theta_q  - \dot \theta_c \phi_q\right]\ ,
\end{eqnarray}
where $S\equiv N(M_0)/2$ is the (dimensionless) spin of the dot. Here $N(M_0)$ is the number of orbital levels of the dot 
in the energy interval $M_0$ around the Fermi energy. Roughly $S = M_0 \bar \rho_{dot}/2$, where 
$\bar \rho_{dot}$ is the density of states averaged over the energy interval $M_0$. The effects of mesoscopic 
fluctuations of the density of states were considered in Ref.~\cite{PhysRevB.85.155311}.  

{\it AES action.} 
The central result of the current paper is the AES-like~\cite{AES_PRL,AES_PRB} effective action, 
which we obtain by expanding 
(\ref{SPhiRotated}) to the first order in $\tilde R^\dag\tilde \Sigma\tilde R$:
$i{\cal S}_{AES} =- \mathrm{tr} \left[\tilde G_{dot,z}
{\tilde R}^{\dag}\,\tilde \Sigma\,\tilde R  \right]$. 
This gives
\begin{eqnarray}\label{SAESMATRIX}
&&i{\cal S}_{AES} =- g \int dt_1 d t_2 \nonumber\\
&&\mathrm{tr} \left[\left(\begin{array}{cc}
R_c^{\dag}(t_1) & \frac{R_q^{\dag}(t_1)}{2}\end{array}
\right)
\left(
\begin{array}{cc}
0 & \alpha_A\\
\alpha_R & \alpha_K
\end{array}
\right)_{(t_1-t_2)}
\left(\begin{array}{c}
R_c(t_2) \\ \frac{R_q(t_2)}{2}\end{array}
\right)
 \right]\ ,\nonumber\\
\end{eqnarray}
where $g = \frac{1}{2}\left(
\pi |V|^2 \rho_{dot}^{\uparrow}\rho_{lead}^{\phantom{\uparrow}} + \pi |V|^2 \rho_{dot}^{\downarrow}\rho_{lead}^{\phantom{\uparrow}} \right)$ is the (spin-independent) conductance per spin direction. 
Here $\rho_{dot}^{\uparrow/\downarrow}$ are the densities of states at the respective $\uparrow$ and $\downarrow$
Fermi levels, whereas the 
density of states in the lead, $\rho_{lead}$, is spin independent.
The standard~\cite{AES_PRB} Ohmic kernel functions are given by
$\alpha_{R}(\omega)- \alpha_{A}(\omega)= 2 \omega$ and $\alpha_{K}(\omega) =2 \omega \coth(\omega/2T)$.
The action (\ref{SAESMATRIX})
strongly resembles the AES action~\cite{AES_PRB}, with $U(1)$ exponents $\exp{\left[i\varphi/2\right]}$ replaced by the $SU(2)$ 
matrices $R$. Fixing the gauge of $R$ is an essential part of our procedure.

{\it Semi-classical equations of motion.} 
From the effective action (\ref{SAESMATRIX}) we derive the following semi-classical equation of motion 
(see~\cite{ASchmid82_Langevin} and Supplemental Material B for details)
\begin{equation}
\begin{split}
&\dot\theta_c +  \tilde g\,
\sin\theta_c \dot \phi_c = \eta_\theta\ ,\\
&\sin\theta_c \left(\dot \phi_c - \gamma B\right)  - \tilde g\, \dot\theta_c  = \eta_\phi \ , 
\end{split}
\label{LLGLangevin}
\end{equation}
where $\tilde g \equiv \frac{g}{2S}$ and  $\gamma=(J\bar \rho_{dot})^{-1}$ is the ``giro-magnetic'' constant of order unity. 
The Langevin forces (torques) are given by 
\begin{eqnarray}
\label{etas}
\eta_\theta =&\phantom{-}&\frac{1}{2S}\, \cos\frac{\theta_c}{2}\,\left[ \xi_x \,\cos\left(\phi_c-\frac{\chi_c}{2}\right) + \xi_y \,\sin\left(\phi_c-\frac{\chi_c}{2}\right)\right]
\nonumber\\
&-&\frac{1}{2S}\, \sin\frac{\theta_c}{2}\,\left[ \xi_z \,\cos\frac{\chi_c}{2} + \xi_0 \,\sin\frac{\chi_c}{2}\right]\ ,\nonumber\\
\eta_\phi  = &-& \frac{1}{2S}\, \cos\frac{\theta_c}{2} \left[\xi_x \,\sin\left(\phi_c-\frac{\chi_c}{2}\right) -
\xi_y \,\cos\left(\phi_c-\frac{\chi_c}{2}\right)\right] \nonumber\\
&-& \frac{1}{2S}\,\sin\frac{\theta_c}{2}\left[\xi_z\,\sin\frac{\chi_c}{2}-\xi_0\,\cos\frac{\chi_c}{2}\right]\ .
\end{eqnarray}
The l.h.s.~of Eqs.~(\ref{LLGLangevin}) represent the standard Landau-Lifshitz-Gilbert (LLG) equations~\cite{Gilbert2004} (without a random torque).  The r.h.s.~represent the random Langevin torque. The latter is expressed in terms of four independent stochastic variables $\xi_j$ ($j=0,x,y,z$), which satisfy  $\langle \xi_j(t_1) \xi_k(t_2)\rangle =  \delta_{jk}\,g\,\alpha_{K}(t_1-t_2)$ and $\langle \xi_j \rangle =0$. 
On the gaussian level, i.e., if fluctuations of $\theta_c$ and $\phi_c$ are neglected in Eqs.~(\ref{etas}), the Langevin forces $\eta_\theta$ 
and $\eta_\phi$ are independent of each other and have the same autocorrelation 
functions:
$\langle \eta_\theta(t_1)\eta_\phi(t_2)\rangle = 0$ and $\langle \eta_\theta(t_1)\eta_\theta(t_2)\rangle = 
\langle \eta_\phi(t_1)\eta_\phi(t_2)\rangle$.
We emphasize that, in general, the noise depends on the angles $\theta_c$ and $\phi_c$ leading to complicated 
dynamics within Eqs.~(\ref{LLGLangevin}). In the classical domain, i.e., for frequencies 
much lower than $T$, we can approximate $\langle \xi_j(t_1) \xi_k(t_2)\rangle = 4gT \delta(t_1 - t_2)\, \delta_{jk}$. 
Then $\langle \eta_\phi(t_1)\eta_\phi(t_2) \rangle =\langle \eta_\theta(t_1)\eta_\theta(t_2) \rangle= (gT/S^2) \delta(t_1 - t_2)$. 
Thus, the situation is simple and we reproduce Ref.~\cite{BrownLLGLangevin}. 

In the quantum high-frequency domain the situation is different. We cannot interpret the four independent 
fields $\xi_n$ as representing the components of a fluctuating magnetic field.  
Solving Eqs.~(\ref{LLGLangevin}) for $\dot \theta$ and $\dot \phi$ we obtain (see Refs.~\cite{BrownLLGLangevin,ChudnovskiyPRL})
\begin{equation}
\begin{split}
&\dot\phi_c - \tilde B  =\frac{1}{\sin\theta_c}\,\xi_\phi\ ,\\
&\dot\theta_c + \sin\theta_c\,\tilde g \tilde B =\xi_\theta\ ,
\end{split}
\label{LLGresolved}
\end{equation} 
where $\xi_\phi \equiv \frac{\eta_\phi + \tilde g \eta_\theta}{1+\tilde g^2}$ and $\xi_\theta \equiv \frac{\eta_\theta - \tilde g \eta_\phi}{1+\tilde g^2}$ and $\tilde B \equiv \frac{\gamma B}{1+\tilde g^2}$. 
A close inspection of these equations shows that in the regime of weak dissipation, $S\gg 1$ and $\tilde g \ll 1$, 
the spin can precess with frequency $\tilde B$ at an almost constant $\theta$ for a long time of order (shorter than) $(\tilde g \tilde B)^{-1}$. For such times we can approximate $\phi_c = \tilde B t$ and 
$\chi_c = (1-\cos\theta_c)\phi_c = (1-\cos\theta_c)\tilde B t$. Thus the Langevin fields $\xi_n$ in (\ref{etas}) are multiplied by 
fast oscillating cosines and sines with frequencies $\omega_c\equiv\tilde B \cos^2(\theta_c/2)$ and $\omega_s\equiv\tilde B \sin^2(\theta_c/2)$.
Thus~\footnote{Here we have dropped non-stationary terms depending
on $t_1+t_2$}
\begin{gather}
\langle \eta_{\phi,\theta}(t_1) \eta_{\phi,\theta}(t_2)\rangle_{\omega=0}=
\frac{g}{4S^2} 
\Bigl [ \cos^2({\theta_c}/{2}) 
\, \alpha_K \left(\omega_c\right)\notag \\
+
\sin^2({\theta_c}/{2}) 
\, \alpha_K \left(\omega_s\right)\Bigr ]\ .
\label{etaeta}
\end{gather}
In the quantum regime $T \ll \tilde B$ these correlation functions differ substantially from the classical ones, 
$\langle \eta_\phi(t) \eta_\phi(t')\rangle_{\omega=0}=\langle \eta_\theta(t) \eta_\theta(t')\rangle_{\omega=0} = g T /S^2$.
Thus, if the spin could be held on a constant $\theta$ trajectory for a long time, the diffusion would be determined 
by the quantum noise at frequencies $\omega_c$ and $\omega_s$, which are governed by the geometric phase.

We are now ready to discuss the physical meaning of the semi-classical approximation, i.e., the expansion of the 
action (\ref{SAESMATRIX}) up to the second order in $\theta_q$ and $\phi_q$ (see also Supplemental Material C). 
The non-expanded action is periodic 
in both $\theta_q$ and $\phi_q$. The periodicity in $\phi_q$ corresponds to the quantization of the $z$ spin component 
$S_z = S \cos\theta_c$. By expanding we restrict ourselves 
to the long time limit, in which $S_z$ has already ''jumped'' many times by $\Delta S_z=1$ in the course of spin diffusion. 
We neglect, thus, higher than the second cumulants of spin noise (see, e.g., Ref.~\cite{Altland2010} for 
similar discussion of charge noise). We obtain, however, a correct second cumulant with down-converted quantum noise 
(similar to shot noise in the charge sector). This is due to the ''multiplicative noise'' character of our Keldysh action (\ref{SAESMATRIX})
similar to the original AES case~\cite{AES_PRB} (see also~\cite{GutmanPRB2005}).

\begin{figure}
 \includegraphics[scale=.3]{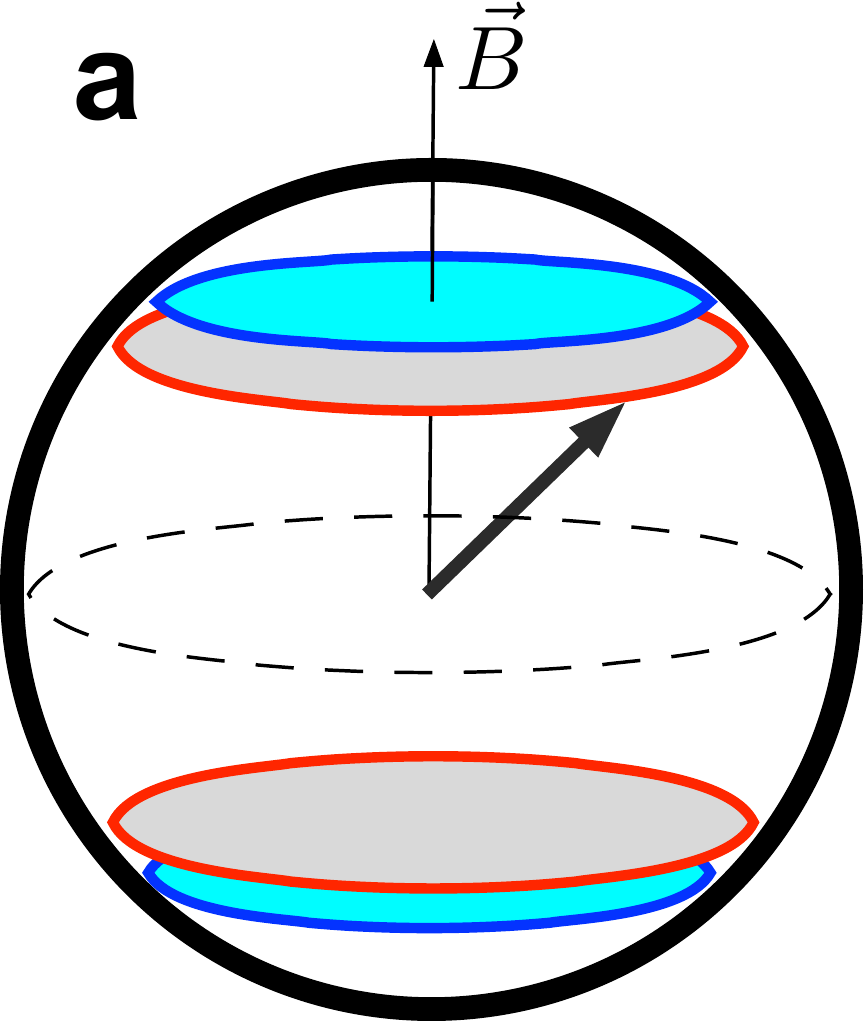}
 \includegraphics[scale=.3]{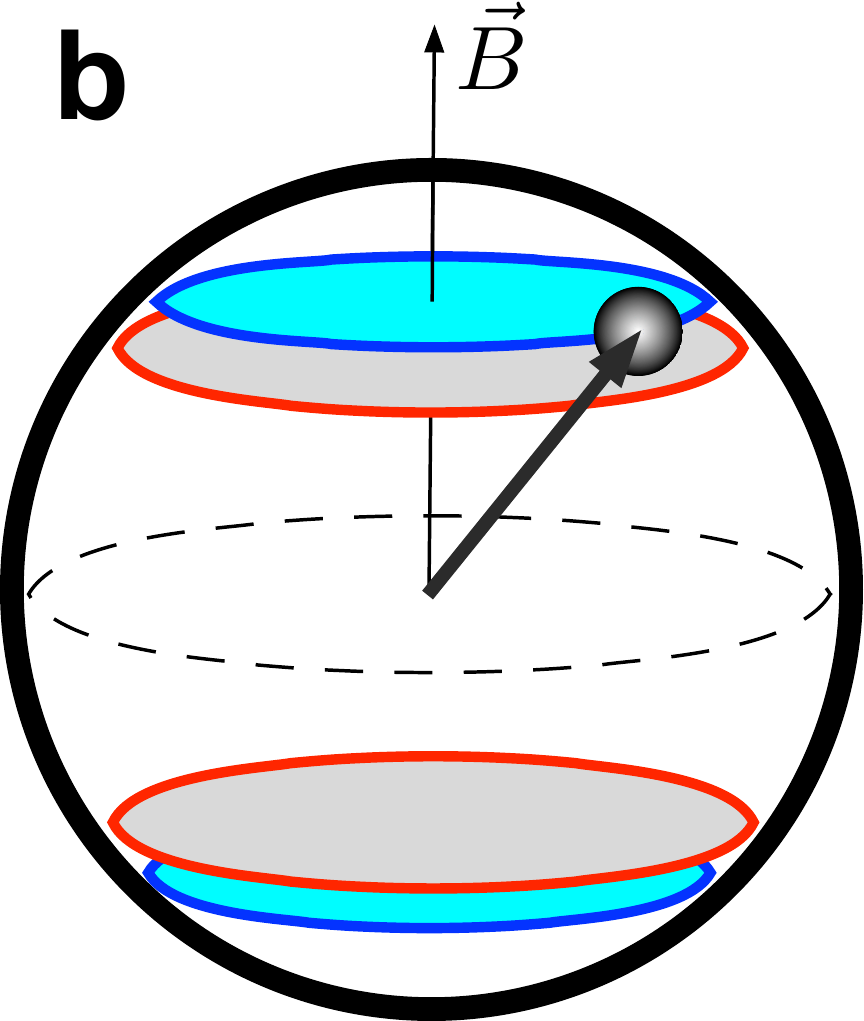}
 \includegraphics[scale=.3]{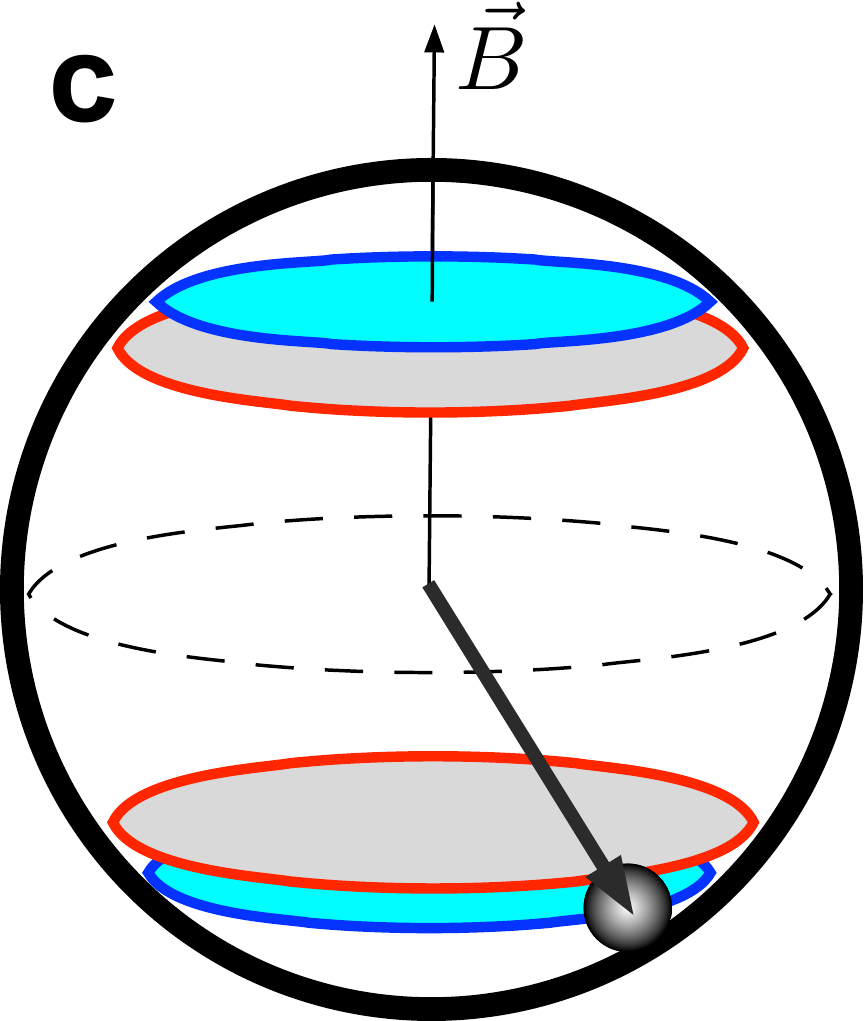}
 \vskip 5mm
 \includegraphics[scale=.3]{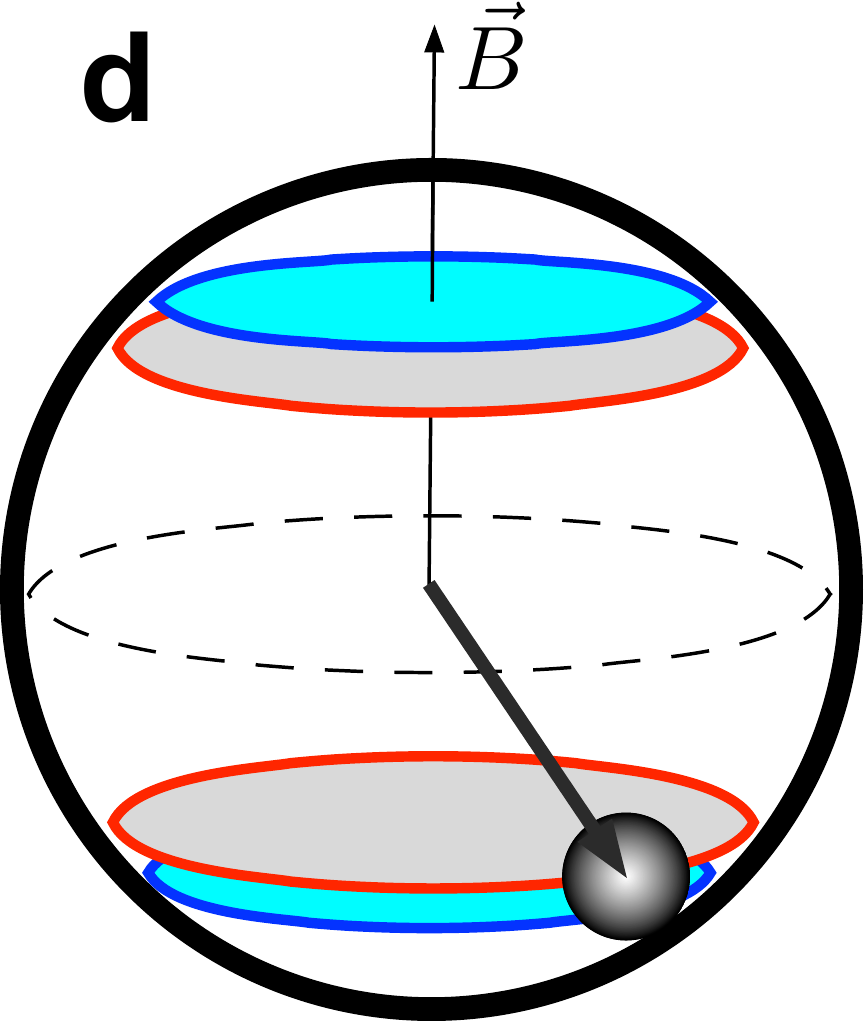}
 \includegraphics[scale=.3]{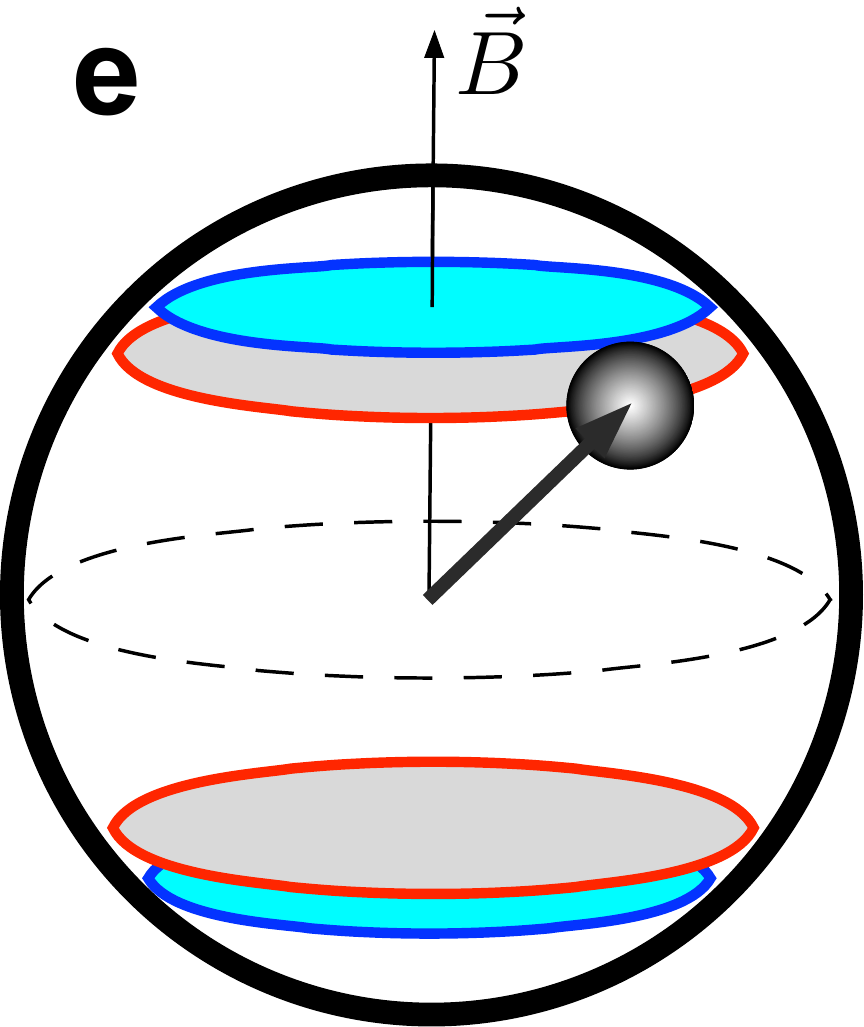}
\caption{''Bang-bang'' protocol. Standard NMR techniques in the rotating frame are employed: 
Rotations around, e.g., the $x$- and the $y$-axes of the rotating frame are achieved by applying resonant driving 
pulses, which are $\pi/2$ phase shifted with respect to each other.  
a) First, a $\theta_0$-pulse around the $y$-axis drives the spin in the $xz$-plane of the rotating frame to form angle 
$\theta_0$ with the $z$-axis;
b) During time $\Delta t \ll \tau_{rel}$ the spin is left alone and it relaxes to $\theta=\theta_0-\delta\theta$, 
where $\delta \theta \approx  \tilde g \tilde B\sin\theta_0\,\Delta t \ll \pi$; 
c) A $\pi$-pulse around $x$ is performed. The spin is again in the $xz$-plane but at $\theta = \pi/2 - (\theta_0 -\delta\theta)$;
d) The spin is left alone again for time $\Delta t$. The relaxation brings it to $\theta = \pi/2 - \theta_0$;
e) A $\pi$-pulse around $x$ is performed. The spin returns to $\theta = \theta_0$ in the $xz$-plane.
This cycle is repeated multiple times. At the end a $-\theta_0$ pulse around $y$-axis would bring the spin back to the north pole, 
but with an accumulated uncertainty (gray cloud in all pannels) due to the quantum geometric diffusion.}
\label{BangBang}
\end{figure}

{\it Measurement protocol.} 
The simplest idea on how to observe the Langevin terms influenced by the Berry phase, would be to perform a Ramsey protocol~\cite{Ramsey} to measure dephasing. Unfortunately this is not a viable option, as for $T\ll \tilde B$ the deterministic relaxation time 
$\tau_{rel}\sim (\tilde g \tilde B)^{-1} \sim S\,(g\tilde B)^{-1}$ is much shorter than the characteristic diffusion time 
$\tau_{diff}\sim S^2\,(g\tilde B)^{-1}$. Thus at the time at which substantial dephasing takes place, the spin is 
long at the north pole ($\theta=0$). To circumvent this hurdle we propose to implement a 
''bang-bang'' protocol~\cite{BangBang} as shown in Fig.~\ref{BangBang} (see also Supplemental Material D). In our protocol we keep the spin at $\theta_c \approx \pm \theta_0$ for a long time. Thus the diffusion will be 
determined by the noise (\ref{etaeta}) at $\theta_c = \theta_0$. 
More precisely, the spread of $\theta_c$ and $\phi_c$ (in the rotating frame) will be given 
by $(\Delta\theta)^2 = \sin^2\theta_0\,(\Delta \varphi)^2 = D t$, where 
\begin{equation}
D = (g/S^2) T_{eff}\ ,
\label{DCoeff}
\end{equation} 
and the effective temperature is calculated from (\ref{etaeta}) to be 
\footnote{Note that very close to the poles these considerations stops working, and $T_{eff}$  (\ref{Teff}) loses its meaning.}
\begin{eqnarray}
T_{eff} &=& \frac{\tilde B}{2}\,\cos^4\left (\frac{\theta_0}{2}\right )\coth\left[\frac{\tilde B}{2T}\cos^2\left (\frac{\theta_0}{2}\right ) \right]
\nonumber\\&+&
\frac{\tilde B}{2}\,\sin^4\left (\frac{\theta_0}{2}\right )\coth\left[\frac{\tilde B}{2T}\sin^2\left (\frac{\theta_0}{2}\right ) \right]\ .
\label{Teff}
\end{eqnarray}
At $T \gg \tilde B$ we obtain $T_{eff}\approx T$, and the geometric effects are completely washed out. We are thus back to the classical regime of~\cite{BrownLLGLangevin}. In the quantum regime, $T \ll \tilde B$, the effective temperature has a characteristic $\theta_0$ dependence $T_{eff} = (1/2) \tilde B\left[\cos^4(\theta_0/2) + \sin^4(\theta_0/2)\right]$ which, if measured, would provide a direct evidence in favor of the geometric noise derived in this paper. 

{\it {Summary and conclusions.}} We have derived an SU(2) generalization of the AES effective action for a large spin. The latter 
gives rise to 
semi-classical LLG-Langevin equations with Langevin torques being influenced by geometric phases. We 
have proposed here a driving protocol that would allow to observe geometric spin-diffusion in the quantum regime. We envision our 
formalism being applied to a broad range of other problems, e.g., easy-axis spin switching, 
the line-width associated with persistent precession in magnetic tunnel junctions~\cite{ChudnovskiyPRL}, 
or transport in arrays of quantum dots~\cite{AltlandArrays}

{\it Acknowledgements.} AS acknowledges the Weston Visiting Professorship at the Weizmann Institute of Science.
ISB acknowledges the support from RFBR (Grant No. 14-02-00333). 
Results Eqs.~(\ref{DCoeff}) and (\ref{Teff}) were obtained with support from 
Russian Science Foundation (Grant No. 14-42-00044).
The work was supported by the GIF, ISF, Swiss NSF, NCCR QIST.

\section{Supplemental Material}

\subsection{A. Choice of the gauge.}
Here we present a detailed justification of the gauge which is presented 
in Eq.~(\ref{KeldyshGaugeFix}).
Ideally we should have chosen a gauge 
that would lead to $Q_\parallel=0$. Seemingly, this might have been achieved with the choice  
$\dot \chi(t) = \dot \phi(t)  \,(1-\cos\theta(t))$ on both branches of the Keldysh contour. This choice, however, 
violates our desired boundary conditions as the integrals over $\dot \chi$ accumulated between $t=-\infty$ and $t=+\infty$ on the upper and on the lower Keldysh branches are different. Such a difference would show up as non-trivial 
boundary conditions on $\chi_q$ at either $t=-\infty$ or $t=+\infty$. In other words, had we 
selected $\dot \chi(t) = \dot \phi(t)  \,(1-\cos\theta(t))$ we should have violated the requirement $\chi_q(t=\pm \infty)=0$. We note, though, that to linear order in the quantum components the condition $\dot \chi(t) = \dot \phi(t)  \,(1-\cos\theta(t))$ yields 
$\dot \chi_q = \dot \phi_q (1-\cos\theta_c) + \theta_q \sin\theta_c\,\dot\phi_c$, leading to 
$\chi_q(t) = \int\limits^t \, dt' \left[\dot \phi_q(t') (1-\cos\theta_c(t')) + \theta_q(t') \sin\theta_c(t')\,\dot\phi_c(t')\right]
=\phi_q(t) (1-\cos\theta_c(t)) +  \int\limits^t \, dt' \,\sin\theta_c(t') \Big[\theta_q(t')\,\dot\phi_c(t') -\dot\theta_c(t')\,\phi_q(t')  \Big]$. The first term vanishes at $t=\pm \infty$ but not the last term. 
We thus include only the first term in $\chi_q$, leading to Eq.~(\ref{KeldyshGaugeFix}), and consequently 
to a non-vanishing contribution to $\tilde Q_\parallel$ (Eq.~(\ref{Qparq})).

\section{B. Semi-classical equations of motion.}

Here we present the derivation of the semiclassical equations of motion, Eq.~(\ref{LLGLangevin}). 
Using the representation 
$R = A_0 \sigma_0 + i A_x \sigma_x + i A_y \sigma_y + i A_z \sigma_z$, with 
$A_0 \equiv  \cos\left[\frac{\theta}{2}\right] \, \cos\left[\frac{\chi}{2}\right]$, 
$A_x \equiv \sin\left[\frac{\theta}{2}\right] \sin{\left[\phi-\frac{\chi}{2}\right]}$,
$A_y \equiv - \sin\left[\frac{\theta}{2}\right] \cos{\left[\phi-\frac{\chi}{2}\right]}$,
$A_z \equiv - \cos\left[\frac{\theta}{2}\right] \, \sin\left[\frac{\chi}{2}\right]$
we rewrite the AES action (Eq.~(\ref{SAESMATRIX})) as ${\cal S}_{AES} ={\cal S}_{AES}^R + {\cal S}_{AES}^K$, where
\begin{eqnarray}\label{SRAES}
i{\cal S}^R_{AES} &=& 
- 2 i g \int dt_1 d t_2 \,\alpha^{''}_{R}(t_1-t_2) \sum_{j} A^q_j(t_1) A^c_j(t_2)\ ,\nonumber\\
\end{eqnarray}
and
\begin{eqnarray}\label{SKAES}
i{\cal S}^K_{AES} &=& -  \frac{g}{2}
 \int dt_1 d t_2 \,\,\alpha_{K}(t_1-t_2)  \sum_{j} A^q_j(t_1) A^q_j(t_2)\ .\nonumber\\
\end{eqnarray}
Here $\alpha_R^{''}(t) \equiv {\rm Im}\,\alpha_R(t)$ and $j=0,x,y,z$.
The Keldysh part of the action (\ref{SKAES}) leads to random Langevin forces. This can be shown~\cite{ASchmid82_Langevin} using 
the Hubbard-Stratonovich transformation 
\begin{eqnarray}
&&e^{i{\cal S}_{AES}^K} =\int \left(\prod_{j=0,x,y,z} D \xi_j\right)  \times\nonumber\\&& 
\exp\left[\int dt \left\{i\sum_{j=0,x,y,z} \xi_j A^q_j\right\} +i{\cal S}_\xi \right]\ ,
\end{eqnarray}
where the action ${\cal S}_\xi$ is given by
\begin{equation}
i{\cal S}_\xi = -\frac{1}{2g}\, \sum_j \int dt_1 dt_2  \left[\alpha_{K}\right]^{-1}_{(t_1-t_2)}  \xi_j(t_1)\xi_j(t_2)\ .
\end{equation}
In other words, 
$\langle \xi_j(t_1) \xi_k(t_2)\rangle =  \delta_{jk}\,g\,\alpha_{K}(t_1-t_2)$ and $\langle \xi_j \rangle =0$.
We obtain the Langevin equations Eq.~(\ref{LLGLangevin}) from $\delta {i\cal S}_{total}/\delta \phi_q(t)=\delta i{\cal S}_{total}/\delta \theta_q(t)=0$, where 
$i {\cal S}_{total} \equiv i{\cal S}_{B}+i{\cal S}_{WZNW} +i{\cal S}^R_{AES} +\int dt\, \sum_j i \xi_j A^q_j$. 
Here $
i{\cal S}_{B} =- i S\gamma\,B\int dt \, \sin\theta_c\,\theta_q$ is the action related to the magnetic field (in $z$-direction). 
Prior to performing the variation of the action, the field $\chi$ is replaced according to the gauge fixing choice (Eq.~(\ref{KeldyshGaugeFix})). Finally, we use 
$\alpha_{R}^{''}(t) = (\partial_t +C) \delta(t)$ (the constant $C$ is important for causality but drops in our calculation) and obtain Eqs.~(\ref{LLGLangevin}).

\section{C. Justification of the semi-classical expansion}

Here we justify why a semiclassical expansion of the action, leading to Eq.~(\ref{LLGLangevin}), is applicable.
It is instructive to rewrite the Berry phase action as $i{\cal S}_{WZNW} =i \,\oint_K dt \,p\, \dot \phi$, 
where $p \equiv S (1-\cos\theta)$.
After the Keldysh rotation this gives
\begin{eqnarray}
\label{WZNWcanonical}
i{\cal S}_{WZNW} &=& i \,\int dt \,\left[ p_c \dot \phi_q + p_q \dot\phi_c\right]\nonumber\\
&=&  i \,\int dt \,\left[ - \phi_q \dot p_c + p_q \dot\phi_c\right]\ ,
\end{eqnarray}
where we used the fact that the quantum component $\phi_q$ must vanish at $t=\pm\infty$.
 In contrast to Eq.~(\ref{WZNW}) we do not yet assume the quantum components to be small, thus, e.g.,
 $p_q = p_u - p_d = -S(\cos\theta_u - \cos\theta_d)$. The rest of the action can in principle be also expressed 
 using these variables. 

The Berry phase part of the action (\ref{WZNWcanonical}) determines the canonical structure of our theory. 
Namely, we can define two pairs of canonically conjugate variables, i.e., $(-p_c, \phi_q)$ and $(\phi_c,p_q)$. Here $-p_c$ and 
$\phi_c$ play the role of canonical coordinates, whereas $\phi_q$ and $p_q$ are their respective conjugate momenta. We are interested in diffusion, i.e., 
noise, of the coordinates $\phi_c$ and $p_c$. The well established way to estimate the latter is to obtain the 
generating function by introducing 
counting source fields (see, e.g., Ref~\cite{Altland2010}). The counting fields shift the conjugate momenta. For example, to calculate the generating function for cumulants of $-p_c$ the Keldysh partition function should be calculated with a shifted 
conjugate momentum $\phi_q \rightarrow \phi_q + \lambda$. The full action, including the dissipative terms, is periodic 
in $\phi_q$. Thus the generating function is periodic in $\lambda$. This corresponds to the quantization of the conjugate coordinate $-p_c$ which is nothing but the classical component 
of $S_z - S$, where $S_z$ is the $z$ projection of the spin. Thus, in all processes described by our AES action $S_z$ changes 
by $\Delta S_z = \pm 1$, as expected for a spin variable.

In this paper we assume $S \gg 1$. Thus, quantized jumps of $S_z$ give rise to very small ($\sim 1/S$) changes of the angle 
$\theta$. 
This allows us to consider the long time limit of continuous diffusion of $\theta$. This limit is well described by a semi-classical 
approximation, in which the action is expanded up to the second order in the quantum components $\theta_q$ and $\phi_q$. 
By performing this expansion we lose all cumulants higher than the second one. In the second cumulant 
(noise) the high frequency quantum noise is mixed (down-converted). This is due to the fact that the expanded Keldysh component of the action (\ref{SKAES}) still contains the classical components $\theta_c$ and $\phi_c$. Thus, the resulting Langevin equation is ''multiplicative'', i.e., the noise terms (Eq.~(\ref{etas})) contain the coordinates $\theta_c$ and $\phi_c$. Similar mechanism led to the shot noise in the original AES case~\cite{AES_PRB} (see also~\cite{GutmanPRB2005}).

The full action is not a periodic function of $p_q$ (it is, of course, periodic as a function of $\theta_{u/d}$). Thus, no quantization corresponds to the second pair of conjugated variables $(\phi_c,p_q)$. 

\section{D. Feasibility of the bang-bang experiment}

Below we argue that the proposed bang-bang experiment is, in fact, within the realm of the present day technology. 
We note that several works dealing with manipulations of qubits did encounter the problem of the 
resolution of the spin state. In particular the “bang-bang” technique has been successfully applied 
(see e.g.,~\cite{Morton2006} (bang-bang in Fulerene qubits);  \cite{Bylander2011} (bang-bang in Josephson qubits)). The spread of the initial spin state may be quantum-limited and could be less than 2\% of a radian (cf.~\cite{Sank2012} on qubit tomography and supplemental material thereof). 
The state may broaden (on the Bloch sphere) through diffusion in the course of its evolution; even if this broadening is tiny, 
it may be resolved following repeated evolutions. 

Let us discuss this in some detail, in the context of our large spin evolution. During the free evolution between two consecutive bang-bang $\pi$-pulses, the geometric diffusion constant is of order $D \sim g B/S^2$, whereas the relaxation rate (the inverse relaxation time) is of order $Γ=\tau^{-1} \sim gB/S$. The time interval $\Delta t$ between consecutive $\pi$-pulses of the bang-bang protocol is chosen so that the angle $\theta$ does not change much due to the deterministic relaxation. In other words, given that 
typically $\theta \sim \pi/4$, we request that $\Delta\theta^{\text{det}} \sim 1/K \ll 1$ (here $K$ is a large integer). We thus choose 
$\Delta t = \tau/K = 1/(\Gamma K)$. After $N$ bang-bang pulses the spread due to the geometric diffusion is of the order of 
$\Delta\theta^{\text{diff}} \sim \sqrt{DN \Delta t}=\sqrt{N/(KS)}$. 
For the latter quantity to be detectable, we require that it is of order $A/S$ ($A \gg 1$), where $1/S$ is the minimal spread corresponding to quantum uncertainty. We assume here that any spread larger than the quantum uncertainty is detectable 
(this can be achieved by averaging over many repetitions of the same bang-bang procedure). This leads to a condition on the minimum number of bang-bang pulses, $N=KA^2/S$. 

Let us  assume for simplicity very strong $\pi$-pulses, i.e. $\Omega>B$, where $\Omega$ is the amplitude of a $\pi$ pulse.  
Then the diffusion constant during the $\pi$-pulses is equal to  $D^{\text{pulse}} \sim g \Omega/S^2$, and the relaxation rate is 
given by $\Gamma^{\text{pulse}} \sim gΩ/S$. The pulse duration is of order $1/\Omega$  (remember we need half a rotation in a 
$\pi$-pulse). The deterministic change of $\theta$ due to and during $N$ pulses (not counting the free evolution between the pulses) is given by 
$$
d\theta_{\text{det}}^{\text{N pulses}}  \sim \Gamma^{\text{pulse}}  N/\Omega=Ng/S\ .
$$
The diffusive spread due to and during N pulses (not counting the free evolution between pulses) is equal to 
$$
d\theta_{\text{diff}}^{\text{N pulses}}  \sim \sqrt{N D^{\text{pulse}}/\Omega} = \sqrt{Ng/S^2}\ .  
$$
Substituting $N=KA^2/S$ we obtain  $d\theta_{\text{det}}^{\text{N pulses}} = F/S$
and $d\theta_{\text{diff}}^{\text{N pulses}} =\sqrt{F}/S$, where $F=gKA^2/S$.   

It is clear that we need to estimate both $d\theta_{\text{det}}^{\text{N pulses}}$
and $d\theta_{\text{diff}}^{\text{N pulses}}$, as neither of them is cancelled by the bang-bang procedure. 
Both errors become of order $1/S$  for $F=1$, that is for 
$g=S/(KA^2)$. Thus, if the tunneling conductance is smaller than this value the error due to the bang-bang pulses is 
smaller than the spread due to the geometric diffusion and therefore unimportant. 

\section{E. Relation to Kondo problem?}

Below we expand the short argument given in the main text 
leading us to conclude that our model is unrelated to the Kondo problem.  
Our quantum dot Hamiltonian (Eq.~(\ref{eq:Hdot})) does not include a charging term, hence no Kondo physics. 
The best way to realize this model is to think of a large quantum dot with negligible charging energy, as was the case, e.g., in Refs.~\cite{ChudnovskiyPRL,BaskoVavilovPRB2009}.
As a result we are neither in a Coulomb valley, nor at a Coulomb peak. 
In this case three different types of fluctuations may take place: (i) Keeping the total $S$ constant, the $S_z$ component may fluctuate; (ii) $S$ itself may fluctuate. We note that in the vicinity of the macroscopic Stoner instability (on either side), the distance in energy between an $S$ and an $S+1$ configuration is much smaller than the level spacing $\delta$ (it is of order $\delta /S$). Once the temperature (or the dot-lead tunneling strength, see below) is larger 
than this energy, such fluctuations in $S$ are facilitated. (iii) Once the temperature is higher than the charging energy (or the tunneling strength becomes larger than the mean level spacing), the Coulomb energy is irrelevant, and fluctuations in the total number of electrons in the dot are allowed. Clearly, fluctuations of either type (ii) or (iii) (or both) take us beyond any Kondo model. 

We note that the dissipative terms in our equations 
of motion are quadratic in the tunneling amplitude (linear in $g$, cf. for example Eq.~(\ref{LLGLangevin})). 
This has also been the case in Refs.~\cite{ChudnovskiyPRL,BaskoVavilovPRB2009}.  
By contrast, cotunneling (facilitating fluctuations of $S_z$ by $1$, changing neither $S$ nor the total charge), 
which would be the building blocks of high-order Kondo screening processes, is second order in $g$, hence
Kondo physics is not present in our analysis.

In passing we note that one standard scenario where the charging energy, even if present, is not important refers to multi-channel leads (not to confuse with multi-channel Kondo). In this scenario each of the channels is weakly coupled to the dot (the tunneling coupling is $|V|^2$), but the sum of all those couplings renders the lead-dot conductance $g>1$. Under these conditions the charging energy is suppressed, but perturbation in $|V|^2$ is allowed (note that the condition for an underdamped motion of the spin implies $g/S \ll 1$; this allows for $g \gg 1$). 

\section{F. Relevance of longitudinal fluctuations of $M$}

In the main text we made an approximation $M(t) = M_0$, thus neglecting completely the longitudinal fluctuations of the magnetization. Here we discuss the effect of the latter and show that it is unimportant as far as our AES dynamics is concerned. 
As shown in our previous works~\cite{JETPLetters.92.179,SahaAnnals,PhysRevB.85.155311}, in the regime of mesoscopic Stoner instability the statistical fluctuations of $M$ in an isolated dot are of the order $\Delta M \sim \sqrt{M_0 T}$.
Close enough to Stoner instability $M_0 \gg T$ and, thus, $\Delta M \ll M_0$. For an isolated dot these are purely statistical 
fluctuations (fluctuations between different ensemble members) since the total spin is a constant of motion there. In an open 
dot considered here dynamical fluctuations of $M$ become possible. One can show that these will be again limited 
by $\Delta M \sim \sqrt{M_0 T} \ll M_0$. In addition these fluctuation are slow (critical slowdown). Thus, the longitudinal 
fluctuations can be safely neglected in an analysis of the spin dynamics on the Bloch sphere. Clearly, in a ferromagnetic dot 
(ferromagnetic side of Stoner) the longitudinal fluctuations are even less noticeable.

\bibliography{paper.bib}

\begin{thebibliography}{29}%
\makeatletter
\providecommand \@ifxundefined [1]{%
 \@ifx{#1\undefined}
}%
\providecommand \@ifnum [1]{%
 \ifnum #1\expandafter \@firstoftwo
 \else \expandafter \@secondoftwo
 \fi
}%
\providecommand \@ifx [1]{%
 \ifx #1\expandafter \@firstoftwo
 \else \expandafter \@secondoftwo
 \fi
}%
\providecommand \natexlab [1]{#1}%
\providecommand \enquote  [1]{``#1''}%
\providecommand \bibnamefont  [1]{#1}%
\providecommand \bibfnamefont [1]{#1}%
\providecommand \citenamefont [1]{#1}%
\providecommand \href@noop [0]{\@secondoftwo}%
\providecommand \href [0]{\begingroup \@sanitize@url \@href}%
\providecommand \@href[1]{\@@startlink{#1}\@@href}%
\providecommand \@@href[1]{\endgroup#1\@@endlink}%
\providecommand \@sanitize@url [0]{\catcode `\\12\catcode `\$12\catcode
  `\&12\catcode `\#12\catcode `\^12\catcode `\_12\catcode `\%12\relax}%
\providecommand \@@startlink[1]{}%
\providecommand \@@endlink[0]{}%
\providecommand \url  [0]{\begingroup\@sanitize@url \@url }%
\providecommand \@url [1]{\endgroup\@href {#1}{\urlprefix }}%
\providecommand \urlprefix  [0]{URL }%
\providecommand \Eprint [0]{\href }%
\providecommand \doibase [0]{http://dx.doi.org/}%
\providecommand \selectlanguage [0]{\@gobble}%
\providecommand \bibinfo  [0]{\@secondoftwo}%
\providecommand \bibfield  [0]{\@secondoftwo}%
\providecommand \translation [1]{[#1]}%
\providecommand \BibitemOpen [0]{}%
\providecommand \bibitemStop [0]{}%
\providecommand \bibitemNoStop [0]{.\EOS\space}%
\providecommand \EOS [0]{\spacefactor3000\relax}%
\providecommand \BibitemShut  [1]{\csname bibitem#1\endcsname}%
\let\auto@bib@innerbib\@empty
\bibitem [{\citenamefont {Kurland}\ \emph {et~al.}(2000)\citenamefont
  {Kurland}, \citenamefont {Aleiner},\ and\ \citenamefont
  {Altshuler}}]{PhysRevB.62.14886}%
  \BibitemOpen
  \bibfield  {author} {\bibinfo {author} {\bibfnamefont {I.~L.}\ \bibnamefont
  {Kurland}}, \bibinfo {author} {\bibfnamefont {I.~L.}\ \bibnamefont
  {Aleiner}}, \ and\ \bibinfo {author} {\bibfnamefont {B.~L.}\ \bibnamefont
  {Altshuler}},\ }\href {\doibase 10.1103/PhysRevB.62.14886} {\bibfield
  {journal} {\bibinfo  {journal} {Phys. Rev. B}\ }\textbf {\bibinfo {volume}
  {62}},\ \bibinfo {pages} {14886} (\bibinfo {year} {2000})}\BibitemShut
  {NoStop}%
\bibitem [{\citenamefont {Kiselev}\ and\ \citenamefont
  {Gefen}(2006)}]{PhysRevLett.96.066805}%
  \BibitemOpen
  \bibfield  {author} {\bibinfo {author} {\bibfnamefont {M.~N.}\ \bibnamefont
  {Kiselev}}\ and\ \bibinfo {author} {\bibfnamefont {Y.}~\bibnamefont
  {Gefen}},\ }\href {\doibase 10.1103/PhysRevLett.96.066805} {\bibfield
  {journal} {\bibinfo  {journal} {Phys. Rev. Lett.}\ }\textbf {\bibinfo
  {volume} {96}},\ \bibinfo {pages} {066805} (\bibinfo {year}
  {2006})}\BibitemShut {NoStop}%
\bibitem [{\citenamefont {Burmistrov}\ \emph {et~al.}(2010)\citenamefont
  {Burmistrov}, \citenamefont {Gefen},\ and\ \citenamefont
  {Kiselev}}]{JETPLetters.92.179}%
  \BibitemOpen
  \bibfield  {author} {\bibinfo {author} {\bibfnamefont {I.}~\bibnamefont
  {Burmistrov}}, \bibinfo {author} {\bibfnamefont {Y.}~\bibnamefont {Gefen}}, \
  and\ \bibinfo {author} {\bibfnamefont {M.}~\bibnamefont {Kiselev}},\ }\href
  {\doibase 10.1134/S0021364010150117} {\bibfield  {journal} {\bibinfo
  {journal} {JETP Letters}\ }\textbf {\bibinfo {volume} {92}},\ \bibinfo
  {pages} {179} (\bibinfo {year} {2010})}\BibitemShut {NoStop}%
\bibitem [{\citenamefont {Saha}\ \emph {et~al.}(2012)\citenamefont {Saha},
  \citenamefont {Gefen}, \citenamefont {Burmistrov}, \citenamefont {Shnirman},\
  and\ \citenamefont {Altland}}]{SahaAnnals}%
  \BibitemOpen
  \bibfield  {author} {\bibinfo {author} {\bibfnamefont {A.}~\bibnamefont
  {Saha}}, \bibinfo {author} {\bibfnamefont {Y.}~\bibnamefont {Gefen}},
  \bibinfo {author} {\bibfnamefont {I.}~\bibnamefont {Burmistrov}}, \bibinfo
  {author} {\bibfnamefont {A.}~\bibnamefont {Shnirman}}, \ and\ \bibinfo
  {author} {\bibfnamefont {A.}~\bibnamefont {Altland}},\ }\href {\doibase
  http://dx.doi.org/10.1016/j.aop.2012.07.013} {\bibfield  {journal} {\bibinfo
  {journal} {Annals of Physics}\ }\textbf {\bibinfo {volume} {327}},\ \bibinfo
  {pages} {2543} (\bibinfo {year} {2012})}\BibitemShut {NoStop}%
\bibitem [{\citenamefont {Burmistrov}\ \emph {et~al.}(2012)\citenamefont
  {Burmistrov}, \citenamefont {Gefen},\ and\ \citenamefont
  {Kiselev}}]{PhysRevB.85.155311}%
  \BibitemOpen
  \bibfield  {author} {\bibinfo {author} {\bibfnamefont {I.~S.}\ \bibnamefont
  {Burmistrov}}, \bibinfo {author} {\bibfnamefont {Y.}~\bibnamefont {Gefen}}, \
  and\ \bibinfo {author} {\bibfnamefont {M.~N.}\ \bibnamefont {Kiselev}},\
  }\href {\doibase 10.1103/PhysRevB.85.155311} {\bibfield  {journal} {\bibinfo
  {journal} {Phys. Rev. B}\ }\textbf {\bibinfo {volume} {85}},\ \bibinfo
  {pages} {155311} (\bibinfo {year} {2012})}\BibitemShut {NoStop}%
\bibitem [{\citenamefont {Viola}\ and\ \citenamefont {Lloyd}(1998)}]{BangBang}%
  \BibitemOpen
  \bibfield  {author} {\bibinfo {author} {\bibfnamefont {L.}~\bibnamefont
  {Viola}}\ and\ \bibinfo {author} {\bibfnamefont {S.}~\bibnamefont {Lloyd}},\
  }\href {\doibase 10.1103/PhysRevA.58.2733} {\bibfield  {journal} {\bibinfo
  {journal} {Phys. Rev. A}\ }\textbf {\bibinfo {volume} {58}},\ \bibinfo
  {pages} {2733} (\bibinfo {year} {1998})}\BibitemShut {NoStop}%
\bibitem [{\citenamefont {Gilbert}(2004)}]{Gilbert2004}%
  \BibitemOpen
  \bibfield  {author} {\bibinfo {author} {\bibfnamefont {T.~L.}\ \bibnamefont
  {Gilbert}},\ }\href {\doibase 10.1109/TMAG.2004.836740} {\bibfield  {journal}
  {\bibinfo  {journal} {IEEE Transactions on Magnetics}\ }\textbf {\bibinfo
  {volume} {40}},\ \bibinfo {pages} {3443} (\bibinfo {year}
  {2004})}\BibitemShut {NoStop}%
\bibitem [{\citenamefont {Brown}(1963)}]{BrownLLGLangevin}%
  \BibitemOpen
  \bibfield  {author} {\bibinfo {author} {\bibfnamefont {W.~F.}\ \bibnamefont
  {Brown}},\ }\href {\doibase 10.1103/PhysRev.130.1677} {\bibfield  {journal}
  {\bibinfo  {journal} {Phys. Rev.}\ }\textbf {\bibinfo {volume} {130}},\
  \bibinfo {pages} {1677} (\bibinfo {year} {1963})}\BibitemShut {NoStop}%
\bibitem [{\citenamefont {Tserkovnyak}\ \emph {et~al.}(2005)\citenamefont
  {Tserkovnyak}, \citenamefont {Brataas}, \citenamefont {Bauer},\ and\
  \citenamefont {Halperin}}]{RevModPhys.77.1375}%
  \BibitemOpen
  \bibfield  {author} {\bibinfo {author} {\bibfnamefont {Y.}~\bibnamefont
  {Tserkovnyak}}, \bibinfo {author} {\bibfnamefont {A.}~\bibnamefont
  {Brataas}}, \bibinfo {author} {\bibfnamefont {G.~E.~W.}\ \bibnamefont
  {Bauer}}, \ and\ \bibinfo {author} {\bibfnamefont {B.~I.}\ \bibnamefont
  {Halperin}},\ }\href {\doibase 10.1103/RevModPhys.77.1375} {\bibfield
  {journal} {\bibinfo  {journal} {Rev. Mod. Phys.}\ }\textbf {\bibinfo {volume}
  {77}},\ \bibinfo {pages} {1375} (\bibinfo {year} {2005})}\BibitemShut
  {NoStop}%
\bibitem [{\citenamefont {Katsura}\ \emph {et~al.}(2006)\citenamefont
  {Katsura}, \citenamefont {Balatsky}, \citenamefont {Nussinov},\ and\
  \citenamefont {Nagaosa}}]{PhysRevB.73.212501}%
  \BibitemOpen
  \bibfield  {author} {\bibinfo {author} {\bibfnamefont {H.}~\bibnamefont
  {Katsura}}, \bibinfo {author} {\bibfnamefont {A.~V.}\ \bibnamefont
  {Balatsky}}, \bibinfo {author} {\bibfnamefont {Z.}~\bibnamefont {Nussinov}},
  \ and\ \bibinfo {author} {\bibfnamefont {N.}~\bibnamefont {Nagaosa}},\ }\href
  {\doibase 10.1103/PhysRevB.73.212501} {\bibfield  {journal} {\bibinfo
  {journal} {Phys. Rev. B}\ }\textbf {\bibinfo {volume} {73}},\ \bibinfo
  {pages} {212501} (\bibinfo {year} {2006})}\BibitemShut {NoStop}%
\bibitem [{\citenamefont {Bode}\ \emph {et~al.}(2012)\citenamefont {Bode},
  \citenamefont {Arrachea}, \citenamefont {Lozano}, \citenamefont {Nunner},\
  and\ \citenamefont {von Oppen}}]{PhysRevB.85.115440}%
  \BibitemOpen
  \bibfield  {author} {\bibinfo {author} {\bibfnamefont {N.}~\bibnamefont
  {Bode}}, \bibinfo {author} {\bibfnamefont {L.}~\bibnamefont {Arrachea}},
  \bibinfo {author} {\bibfnamefont {G.~S.}\ \bibnamefont {Lozano}}, \bibinfo
  {author} {\bibfnamefont {T.~S.}\ \bibnamefont {Nunner}}, \ and\ \bibinfo
  {author} {\bibfnamefont {F.}~\bibnamefont {von Oppen}},\ }\href {\doibase
  10.1103/PhysRevB.85.115440} {\bibfield  {journal} {\bibinfo  {journal} {Phys.
  Rev. B}\ }\textbf {\bibinfo {volume} {85}},\ \bibinfo {pages} {115440}
  (\bibinfo {year} {2012})}\BibitemShut {NoStop}%
\bibitem [{\citenamefont {Chudnovskiy}\ \emph {et~al.}(2008)\citenamefont
  {Chudnovskiy}, \citenamefont {Swiebodzinski},\ and\ \citenamefont
  {Kamenev}}]{ChudnovskiyPRL}%
  \BibitemOpen
  \bibfield  {author} {\bibinfo {author} {\bibfnamefont {A.~L.}\ \bibnamefont
  {Chudnovskiy}}, \bibinfo {author} {\bibfnamefont {J.}~\bibnamefont
  {Swiebodzinski}}, \ and\ \bibinfo {author} {\bibfnamefont {A.}~\bibnamefont
  {Kamenev}},\ }\href {\doibase 10.1103/PhysRevLett.101.066601} {\bibfield
  {journal} {\bibinfo  {journal} {Phys. Rev. Lett.}\ }\textbf {\bibinfo
  {volume} {101}},\ \bibinfo {pages} {066601} (\bibinfo {year}
  {2008})}\BibitemShut {NoStop}%
\bibitem [{\citenamefont {Basko}\ and\ \citenamefont
  {Vavilov}(2009)}]{BaskoVavilovPRB2009}%
  \BibitemOpen
  \bibfield  {author} {\bibinfo {author} {\bibfnamefont {D.~M.}\ \bibnamefont
  {Basko}}\ and\ \bibinfo {author} {\bibfnamefont {M.~G.}\ \bibnamefont
  {Vavilov}},\ }\href {\doibase 10.1103/PhysRevB.79.064418} {\bibfield
  {journal} {\bibinfo  {journal} {Phys. Rev. B}\ }\textbf {\bibinfo {volume}
  {79}},\ \bibinfo {pages} {064418} (\bibinfo {year} {2009})}\BibitemShut
  {NoStop}%
\bibitem [{\citenamefont {Ambegaokar}\ \emph {et~al.}(1982)\citenamefont
  {Ambegaokar}, \citenamefont {Eckern},\ and\ \citenamefont
  {Sch\"on}}]{AES_PRL}%
  \BibitemOpen
  \bibfield  {author} {\bibinfo {author} {\bibfnamefont {V.}~\bibnamefont
  {Ambegaokar}}, \bibinfo {author} {\bibfnamefont {U.}~\bibnamefont {Eckern}},
  \ and\ \bibinfo {author} {\bibfnamefont {G.}~\bibnamefont {Sch\"on}},\ }\href
  {\doibase 10.1103/PhysRevLett.48.1745} {\bibfield  {journal} {\bibinfo
  {journal} {Phys. Rev. Lett.}\ }\textbf {\bibinfo {volume} {48}},\ \bibinfo
  {pages} {1745} (\bibinfo {year} {1982})}\BibitemShut {NoStop}%
\bibitem [{\citenamefont {Eckern}\ \emph {et~al.}(1984)\citenamefont {Eckern},
  \citenamefont {Sch\"on},\ and\ \citenamefont {Ambegaokar}}]{AES_PRB}%
  \BibitemOpen
  \bibfield  {author} {\bibinfo {author} {\bibfnamefont {U.}~\bibnamefont
  {Eckern}}, \bibinfo {author} {\bibfnamefont {G.}~\bibnamefont {Sch\"on}}, \
  and\ \bibinfo {author} {\bibfnamefont {V.}~\bibnamefont {Ambegaokar}},\
  }\href {\doibase 10.1103/PhysRevB.30.6419} {\bibfield  {journal} {\bibinfo
  {journal} {Phys. Rev. B}\ }\textbf {\bibinfo {volume} {30}},\ \bibinfo
  {pages} {6419} (\bibinfo {year} {1984})}\BibitemShut {NoStop}%
\bibitem [{Note1()}]{Note1}%
  \BibitemOpen
  \bibinfo {note} {Here we disregard the charging part of the ''universal''
  Hamiltonian, having in mind, e.g., systems of the type considered in
  Refs.~\cite {ChudnovskiyPRL,BaskoVavilovPRB2009}. Consequently no Kondo
  physics is expected. (See also Supplemental Material E.)}\BibitemShut
  {NoStop}%
\bibitem [{\citenamefont {Kamenev}(2011)}]{KamenevBook}%
  \BibitemOpen
  \bibfield  {author} {\bibinfo {author} {\bibfnamefont {A.}~\bibnamefont
  {Kamenev}},\ }\href@noop {} {\emph {\bibinfo {title} {Field Theory of
  Non-Equilibrium Systems}}}\ (\bibinfo  {publisher} {Cambridge University
  Press, Cambridge},\ \bibinfo {year} {2011})\BibitemShut {NoStop}%
\bibitem [{\citenamefont {Abanov}\ and\ \citenamefont
  {Abanov}(2002)}]{AbanovAbanov}%
  \BibitemOpen
  \bibfield  {author} {\bibinfo {author} {\bibfnamefont {A.~G.}\ \bibnamefont
  {Abanov}}\ and\ \bibinfo {author} {\bibfnamefont {A.}~\bibnamefont
  {Abanov}},\ }\href {\doibase 10.1103/PhysRevB.65.184407} {\bibfield
  {journal} {\bibinfo  {journal} {Phys. Rev. B}\ }\textbf {\bibinfo {volume}
  {65}},\ \bibinfo {pages} {184407} (\bibinfo {year} {2002})}\BibitemShut
  {NoStop}%
\bibitem [{\citenamefont {Volovik}(1987)}]{Volovik87}%
  \BibitemOpen
  \bibfield  {author} {\bibinfo {author} {\bibfnamefont {G.~E.}\ \bibnamefont
  {Volovik}},\ }\href {http://stacks.iop.org/0022-3719/20/i=7/a=003} {\bibfield
   {journal} {\bibinfo  {journal} {Journal of Physics C: Solid State Physics}\
  }\textbf {\bibinfo {volume} {20}},\ \bibinfo {pages} {L83} (\bibinfo {year}
  {1987})}\BibitemShut {NoStop}%
\bibitem [{\citenamefont {Schmid}(1982)}]{ASchmid82_Langevin}%
  \BibitemOpen
  \bibfield  {author} {\bibinfo {author} {\bibfnamefont {A.}~\bibnamefont
  {Schmid}},\ }\href {\doibase 10.1007/BF00681904} {\bibfield  {journal}
  {\bibinfo  {journal} {J. of Low Temp. Phys.}\ }\textbf {\bibinfo {volume}
  {49}},\ \bibinfo {pages} {609} (\bibinfo {year} {1982})}\BibitemShut
  {NoStop}%
\bibitem [{Note2()}]{Note2}%
  \BibitemOpen
  \bibinfo {note} {Here we have dropped non-stationary terms depending on
  $t_1+t_2$}\BibitemShut {NoStop}%
\bibitem [{\citenamefont {Altland}\ \emph {et~al.}(2010)\citenamefont
  {Altland}, \citenamefont {De~Martino}, \citenamefont {Egger},\ and\
  \citenamefont {Narozhny}}]{Altland2010}%
  \BibitemOpen
  \bibfield  {author} {\bibinfo {author} {\bibfnamefont {A.}~\bibnamefont
  {Altland}}, \bibinfo {author} {\bibfnamefont {A.}~\bibnamefont {De~Martino}},
  \bibinfo {author} {\bibfnamefont {R.}~\bibnamefont {Egger}}, \ and\ \bibinfo
  {author} {\bibfnamefont {B.}~\bibnamefont {Narozhny}},\ }\href {\doibase
  10.1103/PhysRevB.82.115323} {\bibfield  {journal} {\bibinfo  {journal} {Phys.
  Rev. B}\ }\textbf {\bibinfo {volume} {82}},\ \bibinfo {pages} {115323}
  (\bibinfo {year} {2010})}\BibitemShut {NoStop}%
\bibitem [{\citenamefont {Gutman}\ \emph {et~al.}(2005)\citenamefont {Gutman},
  \citenamefont {Mirlin},\ and\ \citenamefont {Gefen}}]{GutmanPRB2005}%
  \BibitemOpen
  \bibfield  {author} {\bibinfo {author} {\bibfnamefont {D.~B.}\ \bibnamefont
  {Gutman}}, \bibinfo {author} {\bibfnamefont {A.~D.}\ \bibnamefont {Mirlin}},
  \ and\ \bibinfo {author} {\bibfnamefont {Y.}~\bibnamefont {Gefen}},\ }\href
  {\doibase 10.1103/PhysRevB.71.085118} {\bibfield  {journal} {\bibinfo
  {journal} {Phys. Rev. B}\ }\textbf {\bibinfo {volume} {71}},\ \bibinfo
  {pages} {085118} (\bibinfo {year} {2005})}\BibitemShut {NoStop}%
\bibitem [{\citenamefont {Ramsey}(1950)}]{Ramsey}%
  \BibitemOpen
  \bibfield  {author} {\bibinfo {author} {\bibfnamefont {N.~F.}\ \bibnamefont
  {Ramsey}},\ }\href {\doibase 10.1103/PhysRev.78.695} {\bibfield  {journal}
  {\bibinfo  {journal} {Phys. Rev.}\ }\textbf {\bibinfo {volume} {78}},\
  \bibinfo {pages} {695} (\bibinfo {year} {1950})}\BibitemShut {NoStop}%
\bibitem [{Note3()}]{Note3}%
  \BibitemOpen
  \bibinfo {note} {Note that very close to the poles these considerations stops
  working, and $T_{eff}$ (\ref {Teff}) loses its meaning.}\BibitemShut {Stop}%
\bibitem [{\citenamefont {Altland}\ \emph {et~al.}(2006)\citenamefont
  {Altland}, \citenamefont {Glazman}, \citenamefont {Kamenev},\ and\
  \citenamefont {Meyer}}]{AltlandArrays}%
  \BibitemOpen
  \bibfield  {author} {\bibinfo {author} {\bibfnamefont {A.}~\bibnamefont
  {Altland}}, \bibinfo {author} {\bibfnamefont {L.}~\bibnamefont {Glazman}},
  \bibinfo {author} {\bibfnamefont {A.}~\bibnamefont {Kamenev}}, \ and\
  \bibinfo {author} {\bibfnamefont {J.}~\bibnamefont {Meyer}},\ }\href
  {\doibase http://dx.doi.org/10.1016/j.aop.2005.12.012} {\bibfield  {journal}
  {\bibinfo  {journal} {Annals of Physics}\ }\textbf {\bibinfo {volume}
  {321}},\ \bibinfo {pages} {2566 } (\bibinfo {year} {2006})}\BibitemShut
  {NoStop}%
\bibitem [{\citenamefont {Morton}\ \emph {et~al.}(2006)\citenamefont {Morton},
  \citenamefont {Tyryshkin}, \citenamefont {Ardavan}, \citenamefont {Benjamin},
  \citenamefont {Porfyrakis}, \citenamefont {Lyon},\ and\ \citenamefont
  {Briggs}}]{Morton2006}%
  \BibitemOpen
  \bibfield  {author} {\bibinfo {author} {\bibfnamefont {J.~J.~L.}\
  \bibnamefont {Morton}}, \bibinfo {author} {\bibfnamefont {A.~M.}\
  \bibnamefont {Tyryshkin}}, \bibinfo {author} {\bibfnamefont {A.}~\bibnamefont
  {Ardavan}}, \bibinfo {author} {\bibfnamefont {S.~C.}\ \bibnamefont
  {Benjamin}}, \bibinfo {author} {\bibfnamefont {K.}~\bibnamefont
  {Porfyrakis}}, \bibinfo {author} {\bibfnamefont {S.~A.}\ \bibnamefont
  {Lyon}}, \ and\ \bibinfo {author} {\bibfnamefont {G.~A.~D.}\ \bibnamefont
  {Briggs}},\ }\href {http://dx.doi.org/10.1038/nphys192} {\bibfield  {journal}
  {\bibinfo  {journal} {Nature Physics}\ }\textbf {\bibinfo {volume} {2}},\
  \bibinfo {pages} {40} (\bibinfo {year} {2006})}\BibitemShut {NoStop}%
\bibitem [{\citenamefont {Bylander}\ \emph {et~al.}(2011)\citenamefont
  {Bylander}, \citenamefont {Gustavsson}, \citenamefont {Yan}, \citenamefont
  {Yoshihara}, \citenamefont {Harrabi}, \citenamefont {Fitch}, \citenamefont
  {Cory}, \citenamefont {Nakamura}, \citenamefont {Tsai},\ and\ \citenamefont
  {Oliver}}]{Bylander2011}%
  \BibitemOpen
  \bibfield  {author} {\bibinfo {author} {\bibfnamefont {J.}~\bibnamefont
  {Bylander}}, \bibinfo {author} {\bibfnamefont {S.}~\bibnamefont
  {Gustavsson}}, \bibinfo {author} {\bibfnamefont {F.}~\bibnamefont {Yan}},
  \bibinfo {author} {\bibfnamefont {F.}~\bibnamefont {Yoshihara}}, \bibinfo
  {author} {\bibfnamefont {K.}~\bibnamefont {Harrabi}}, \bibinfo {author}
  {\bibfnamefont {G.}~\bibnamefont {Fitch}}, \bibinfo {author} {\bibfnamefont
  {D.~G.}\ \bibnamefont {Cory}}, \bibinfo {author} {\bibfnamefont
  {Y.}~\bibnamefont {Nakamura}}, \bibinfo {author} {\bibfnamefont {J.-S.}\
  \bibnamefont {Tsai}}, \ and\ \bibinfo {author} {\bibfnamefont {W.~D.}\
  \bibnamefont {Oliver}},\ }\href {http://dx.doi.org/10.1038/nphys1994}
  {\bibfield  {journal} {\bibinfo  {journal} {Nature Physics}\ }\textbf
  {\bibinfo {volume} {7}},\ \bibinfo {pages} {565} (\bibinfo {year}
  {2011})}\BibitemShut {NoStop}%
\bibitem [{\citenamefont {Sank}\ \emph {et~al.}(2012)\citenamefont {Sank},
  \citenamefont {Barends}, \citenamefont {Bialczak}, \citenamefont {Chen},
  \citenamefont {Kelly}, \citenamefont {Lenander}, \citenamefont {Lucero},
  \citenamefont {Mariantoni}, \citenamefont {Megrant}, \citenamefont {Neeley},
  \citenamefont {O'Malley}, \citenamefont {Vainsencher}, \citenamefont {Wang},
  \citenamefont {Wenner}, \citenamefont {White}, \citenamefont {Yamamoto},
  \citenamefont {Yin}, \citenamefont {Cleland},\ and\ \citenamefont
  {Martinis}}]{Sank2012}%
  \BibitemOpen
  \bibfield  {author} {\bibinfo {author} {\bibfnamefont {D.}~\bibnamefont
  {Sank}}, \bibinfo {author} {\bibfnamefont {R.}~\bibnamefont {Barends}},
  \bibinfo {author} {\bibfnamefont {R.~C.}\ \bibnamefont {Bialczak}}, \bibinfo
  {author} {\bibfnamefont {Y.}~\bibnamefont {Chen}}, \bibinfo {author}
  {\bibfnamefont {J.}~\bibnamefont {Kelly}}, \bibinfo {author} {\bibfnamefont
  {M.}~\bibnamefont {Lenander}}, \bibinfo {author} {\bibfnamefont
  {E.}~\bibnamefont {Lucero}}, \bibinfo {author} {\bibfnamefont
  {M.}~\bibnamefont {Mariantoni}}, \bibinfo {author} {\bibfnamefont
  {A.}~\bibnamefont {Megrant}}, \bibinfo {author} {\bibfnamefont
  {M.}~\bibnamefont {Neeley}}, \bibinfo {author} {\bibfnamefont {P.~J.~J.}\
  \bibnamefont {O'Malley}}, \bibinfo {author} {\bibfnamefont {A.}~\bibnamefont
  {Vainsencher}}, \bibinfo {author} {\bibfnamefont {H.}~\bibnamefont {Wang}},
  \bibinfo {author} {\bibfnamefont {J.}~\bibnamefont {Wenner}}, \bibinfo
  {author} {\bibfnamefont {T.~C.}\ \bibnamefont {White}}, \bibinfo {author}
  {\bibfnamefont {T.}~\bibnamefont {Yamamoto}}, \bibinfo {author}
  {\bibfnamefont {Y.}~\bibnamefont {Yin}}, \bibinfo {author} {\bibfnamefont
  {A.~N.}\ \bibnamefont {Cleland}}, \ and\ \bibinfo {author} {\bibfnamefont
  {J.~M.}\ \bibnamefont {Martinis}},\ }\href@noop {} {\bibfield  {journal}
  {\bibinfo  {journal} {Phys. Rev. Lett.}\ }\textbf {\bibinfo {volume} {109}},\
  \bibinfo {pages} {067001} (\bibinfo {year} {2012})}\BibitemShut {NoStop}%
\end{thebibliography}%
\end{document}